\documentclass[12pt,preprint]{aastex}
\usepackage[dvips]{color}

\renewcommand{\d}{\mbox{d}}
\newcommand{\dpp}[2]{\frac{\partial #1}{\partial #2}}

\newcommand{\mi}[1]{\mbox{\boldmath$#1$}}
\newcommand{\Nbv}{\mathcal{N}^{\,2}}

\shorttitle{Influence of Wavefield Strength on Travel Time}
\shortauthors{Parchevsky, Zhao \& Kosovichev}

\begin{document}
\title{Influence of Non-Uniform Distribution of Acoustic Wavefield
Strength on Time-Distance Helioseismology Measurements}

\author{Konstantin V. Parchevsky, Junwei Zhao, and Alexander G. Kosovichev}
\affil{W.~W.~Hansen Experimental Physics Laboratory, Stanford
University, Stanford, CA 94305-4085}
\email{kparchevsky@solar.stanford.edu}

\begin{abstract}
By analyzing numerically simulated solar oscillation data, we study
the influence of non-uniform distribution of acoustic wave
amplitude, acoustic source strength, and perturbations of the sound
speed on the shifts of acoustic travel times measured by the
time-distance helioseismology method. It is found that for short
distances, the contribution to the mean travel time shift caused by
non-uniform distribution of acoustic sources in sunspots may be
comparable to (but smaller than) the contribution from the sound
speed perturbation in sunspots, and that it has the opposite sign to
the sound-speed effect. This effect may cause some underestimation
of the negative sound-speed perturbations in sunspots just below the
surface, that was found in previous time-distance helioseismology
inferences. This effect cannot be corrected by artificially
increasing the amplitude of oscillations in sunspots. For large
time-distance annuli, the non-uniform distribution of wavefields
does not have significant effects on the mean travel times, and thus
the sound-speed inversion results. The measured travel time
differences, which are used to determine the mass flows beneath
sunspots, can also be systematically shifted by this effect, but
only in an insignificant magnitude.
\end{abstract}


\section{Introduction}
Time-distance helioseismology is based on measuring and inverting
acoustic wave travel times between separate points on the surface of
the Sun. It is one of widely used approaches of local
helioseismology for reconstructing solar subsurface structures and
flows. Calculation of the temporal cross-covariance of two
oscillation signals, observed in different points on the solar
surface, is a key element of this method \citep{Duvall1993}.
\citet{Kosovichev1997} showed that the cross-covariance function for
waves with the phase speed lying in a narrow interval can be
approximately represented by a Gabor wavelet. The phase and group
travel times of acoustic waves can be obtained by fitting the Gabor
wavelet to the observed cross-covariance function, using a
least-square method. The measured phase travel times are used for
inferring the subphotospheric perturbations of the sound (wave)
speed and flow velocities in the quiet Sun and sunspot regions
\citep{Kosovichev2000,Zhao2001}. The mean travel time of acoustic
waves traveling between two points in the opposite directions are
used for determining the sound speed, and the travel time difference
is used for determining the flows.

 However, an accurate inference of sunspot's subsurface
sound-speed structures and flow fields by use of this approach may
be affected by a series of physical and unphysical effects, such as
strong wave damping in active regions \citep{woodard97}, and the
presence of strong magnetic field \citep{zhao06}. Recently, by using
observations of a quiet Sun region and artificially reducing solar
acoustic oscillation amplitudes, i.e. masking solar wavefield to
mimic the sunspot's behavior, \citet{Rajaguru2006} found that this
procedure could shift the measured acoustic travel times
systematically by an amount of $5-40\%$, although such a shift was
not expected. Furthermore, they suggested to correct the observed
acoustic wavefields inside active regions by artificially increasing
the wave amplitude.

However, it is evident that the artificially masked wavefield of a
solar quiet region can only mimic the acoustic power of the active
region, but not the actual physical cause. Therefore, the systematic
errors estimated by this approach may be inaccurate, and the
correction procedure is unjustified. In this paper, we have carried
out 3D numerical simulations of solar oscillations based on three
different models to mimic the sunspot's wavefield, and investigated
the systematic errors caused by the amplitude effects in the
time-distance measurements. These models include the artificial
masking the numerically simulated wavefields, as suggested by
\citet{Rajaguru2006}, reducing the strength of oscillation sources
to reflect the physical effect of reduced excitation in sunspots
\citep{Parchevsky2007a}, and compare these with effects caused by
using a sound-speed perturbation deduced by the previous sound-speed
inversions \citep{Kosovichev2000}. The numerical simulation
procedure and results are described in \S2, and the results of the
time-distance analysis are given in \S3, followed by discussions in
\S4.

\section{Numerical Simulations}
\subsection{Governing Equations} Propagation of acoustic waves on the Sun is
described by the system of the linearized Euler equations
\begin{equation}\label{Eq:LinEuler}
\begin{array}{l}
\displaystyle\dpp{\rho'}{t} + \mi{\nabla}\cdot(\rho_0\mi{u}')= 0
\\[9pt]
\displaystyle\dpp{}{t}(\rho_0 \mi{u}') + \mi{\nabla}\;p' = \mi{g}_0
\rho' + \mi{f}(x,y,z,t),
\end{array}
\end{equation}
where $\mi{u}'$ is the velocity perturbation, $\rho'$ and $p'$ are
the density and pressure perturbations respectively, and
$\mi{f}(x,y,z,t)$ is the function describing the acoustic sources.
The pressure $p_0$, density $\rho_0$, and gravitational
accelerations $\mi{g}_0$ with subscripts 0 correspond to the
background model. To close the system (\ref{Eq:LinEuler}) we used
the adiabatic relation $\delta p/p_0 = \Gamma_1 \delta\rho/\rho_0$
between Lagrangian variations of pressure $\delta p$ and density
$\delta\rho$. The adiabatic exponent $\Gamma_1$ is calculated from
the realistic OPAL equation of state \citep{Rogers1996} for the
hydrogen $X$ and heavy elements $Z$ abundances of the standard
model. The standard solar model S \citep{Christensen-Dalsgaard1996}
with a smoothly joined model of the chromosphere of
\citet{Vernazza1976} is used as the background model.

The standard solar model is convectively unstable, especially in the
superadiabatic subphotospheric layers where convective motions are
very intense and turbulent. Using this convectively unstable model
as a background model leads to the instability of the solution of
the linear system. The condition for stability against convection
requires that the square of Brunt-V\"{a}is\"{a}l\"{a} frequency
$\Nbv_0=g_0(1/\Gamma_1\;\d\log p_0/\d r - \d\log \rho_0/\d r)$ is
positive. To make the background model convectively stable we
replaced all negative values of $\Nbv_0$ by zeros and recalculated
the profiles of pressure and density from the condition of
hydrostatic equilibrium. This procedure guarantees convective
stability of the background model. It has been shown that the
profiles of pressure $p_0$, density $\rho_0$, sound speed $c_0$, and
acoustic cut-off frequency $\omega_c^2 = c_0^2/4H_{\rho}^2(1-2\;\d
H_{\rho}/\d r)$ of the modified model are very close to the
corresponding profiles of the standard solar model
\citep{Parchevsky2007b}. Quantity $H_{\rho}^{-1}=-\d \log\rho_0/\d
r$ represents the density hight scale.

To prevent spurious reflections of acoustic waves from the
boundaries we established non-reflecting boundary conditions based
on the Perfectly Matched Layer (PML) method \citep{Hu1996} at the
top and bottom boundaries. The top boundary was set at the height of
500 km above the photosphere. This simulates a realistic situation
when not all waves are reflected from the photosphere. Waves with
frequencies higher than the acoustic cut-off frequency pass through
the photosphere and are absorbed by the top boundary. This naturally
introduces frequency dependence of the reflecting coefficient of the
top boundary. The lateral boundary conditions are periodic. The
details are described by \citet{Parchevsky2007b}.

The waves are generated by spatially localized sources of the
z-component of force
\begin{equation}\label{Eq:FzSource}
\mi{f}(x,y,z,t)=\left\{
\begin{array}{ll}
\displaystyle\mi{e}_zA\left[1-\left(\frac{r}{R_{src}}\right)^2\right]^2
(1-2\tau^2)e^{-\tau^2} & \mbox{if}\quad r\leq R_{src}\\
\displaystyle 0 & \mbox{if}\quad r>R_{src}
\end{array}
\right.
\end{equation}
with $r$ and $\tau$ given by
\begin{equation}
r=\sqrt{(x-x_{src})^2+(y-y_{src})^2+(z-z_{src})^2}, \quad
\tau=\frac{\omega (t-t_0)}{2}-\pi,\quad t_0\leq t\leq
t_0+\frac{4\pi}{\omega}
\end{equation}
where $\mi{e}_z$ is the unit vector in the vertical direction,
$x_{src}$, $y_{src}$, and $z_{src}$ are the coordinates of the
center of the source, $R_{src}$ is the source radius, $\omega$ is
the central frequency, $t_0$ is the moment of the source ignition,
$A$ is the coefficient, which is measured in dyn cm$^{-3}$ and
describes the source strength. It has a physical meaning of the
force density.

\subsection{Numerical Method} To solve the system (\ref{Eq:LinEuler}),
a semi-discrete code developed by \citet{Parchevsky2007b} is used.
The high-order dispersion relation preserving (DRP) finite
difference (FD) scheme of \citet{Tam1993} is used for spatial
discretization. The coefficients of this FD scheme are chosen from
the requirement that the error in the Fourier transform of the
spatial derivative is minimal. It can be shown that the 4th-order
DRP FD scheme describes short waves more accurately than the classic
6th-order FD scheme. A 3rd-order, three-stage strong stability
preserving Runge-Kutta scheme with the Courant number, $C=1$,
\citep{Shu2002} is used as a time advancing scheme.

The efficiency of the high-order FD schemes can be reached only if
they are combined with adequate numerical boundary conditions. We
followed \citet{Carpenter1993} and used an implicit Pad\'{e}
approximation of the spatial derivatives near the top and bottom
boundaries to derive a stable 3rd-order numerical boundary
conditions consistent with the 4th-order DRP numerical scheme for
interior points of the computational domain.

Waves with the wavelength less than $4\Delta x$ are not resolved by
the FD scheme. They lead to point-to-point oscillations of the
solution that can cause a numerical instability. Such waves have to
be filtered out, and we use a 6th-order digital filter to eliminate
unresolved short wave component from the solution at each time step.

\subsection{Simulation of Artificial Data}
The simulations are carried out in a rectangular domain of size
90$\times$90$\times$31 Mm$^3$ using a uniform
600$\times$600$\times$619 grid. The background model varies sharply
in the region above the temperature minimum. Thus, to simulate the
propagation of acoustic waves into the chromosphere we choose the
vertical spatial step $\Delta z =50$ km in order to preserve the
accuracy and numerical stability. The spatial intervals in the
horizontal direction are $\Delta x=\Delta y=150$ km. To satisfy the
Courant stability condition for the explicit scheme, the time step
is set to be equal to 0.5 seconds. The sources of the z-component of
force with random amplitudes and uniform frequency distribution in
the range of $2 - 8$ mHz are randomly distributed at the depth of
100 km below the photosphere and are independently excited at
arbitrary moments of time.

We describe three sets of simulations with different distribution of
acoustic sources and different background models. The first
reference model (Model~I) represents simulations of the acoustic
wave field for horizontally uniform distribution of the acoustic
sources and the horizontally uniform background model. This model
corresponds to the quiet Sun and will be used as a reference state
for the following time-distance analysis. The acoustic travel times
for models II and III are computed relatively to this reference
model. The goal of this study is to estimate the contributions to
the travel times arising from perturbations of the background model
and non-uniform distribution of the acoustic sources separately. For
this purpose, in  model~II the acoustic source strength is gradually
decreased (masked) in the central region,  simulating the reduction
of the acoustic sources in sunspots \citep{Parchevsky2007a}. In this
model, the horizontal axially symmetric distribution of the acoustic
source strength is given by formula
\begin{equation}\label{Eq:SrcMask}
A(x,y)=\left\{
\begin{array}{ll}
\displaystyle \frac{1}{2}\left(1-\cos\frac{\pi r_h}{R_s}\right) &
\mbox{if}\quad r_h\leq R_s\\
\displaystyle 1 & \mbox{if}\quad r_h>R_s,
\end{array}
\right.
\end{equation}
where $r_h=\sqrt{(x-x_s)^2+(y-y_s)^2}$ is the horizontal distance
from the sunspot axis, $x_s$, $y_s$, and $R_s$ are the x-, y-
coordinates of the sunspot center and the sunspot radius
respectively. The background model remains unperturbed and
horizontally uniform. So, as far as the background model remains
unchanged all deviations of simulated wave field properties from
model~I can be explained as a result on non-uniform distribution of
the acoustic sources.

Strictly speaking, travel times for the cases of uniform
distribution of the acoustic sources (Model~I) and masked source
strength (Model~II) are calculated at different conditions. The
amplitude of the wave field is uniform in the first case and
non-uniform in the second one. To take this into account we mask the
{\it wave field} of the Model~I by masking function computed by
averaging signals azimuthally around the sunspot center. This mimics
the reduced amplitude of active regions, just as what was done by
\citet{Rajaguru2006}. The resultant model is called Model~Ia for the
convenience of reference in the following descriptions. Although the
amplitude distributions now are the same for both wave fields, the
wave fields itself are different, because masking of the source
strength is not reduced to simple masking of the resulting wave
field.

\begin{figure}
\plotone{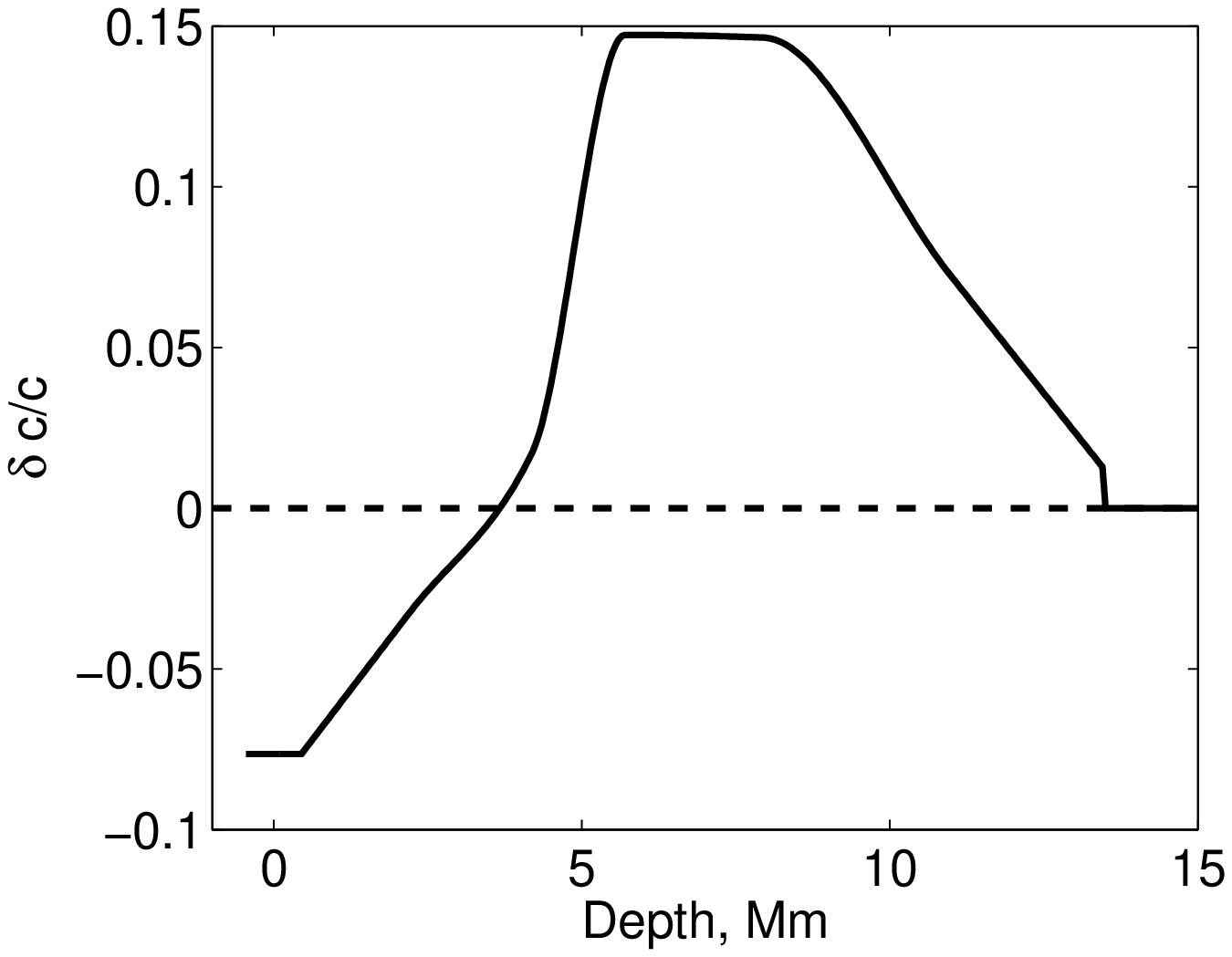} \caption{Perturbation of the sound speed profile at
the sunspot axis, calculated by inverting data obtained from
SOHO/MDI observations of the sunspot on 20 June 1998.} \label{fg1}
\end{figure}

Model~III combines the source masking of Model~II with the sound
speed perturbation in sunspots. The 3D sound-speed profile $c$ in
Model~III is approximated by the formula
\begin{equation}\label{Eq:SndSpeed}
c(x,y,z) = c_0(z)\left[1+\frac{\delta
c(z)}{c_0(z)}(1-A(x,y))\right],
\end{equation}
where $\delta c/c_0$ is the vertical profile of the sound speed
perturbation at the sunspot axis. This profile is shown in
Figure~\ref{fg1} and was calculated from the inversion of
helioseismic data for the sunspot observed by SOHO/MDI on 20 June
1998 \citep{Kosovichev2000}. The change of the sound-speed
perturbation sign from negative to positive at approximately 4 Mm
below the photosphere  is a characteristic feature of this profile.
The depth of inversion divides the domain in vertical direction into
two regions with the sound speed greater and smaller than in the
standard reference model. Hence, we expect different behavior of
waves propagating though this artificial sunspot if their turning
points lie in different regions.

\begin{figure}
\plotone{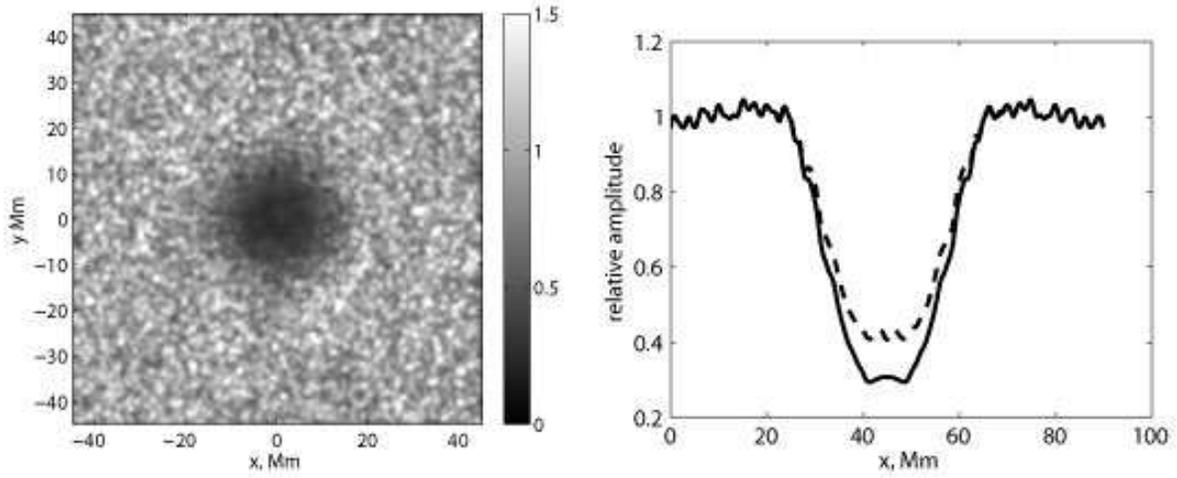} \caption{Amplitude map ({\it left}) of the vertical
velocity component for simulations with masked sources and sound
speed perturbation of the background model. The solid line
represents azimuthally averaged profile of the map. The dashed line
shows azimuthally averaged profile of the vertical velocity
component for simulations with masked sources only (the background
model is horizontally uniform).} \label{fg2}
\end{figure}

The amplitude map of the resulting wave field for Model~III is shown
in the left panel of Figure~\ref{fg2}. The solid line in the right
panel represents the azimuthally averaged amplitude profile. The
dashed line shows the angularly averaged amplitude profile for
Model~II. The inhomogeneity of the sound speed causes increasing the
ratio of oscillation amplitudes outside and inside of the artificial
sunspot by about 40\%.

\begin{figure}
\epsscale{0.7} \plotone{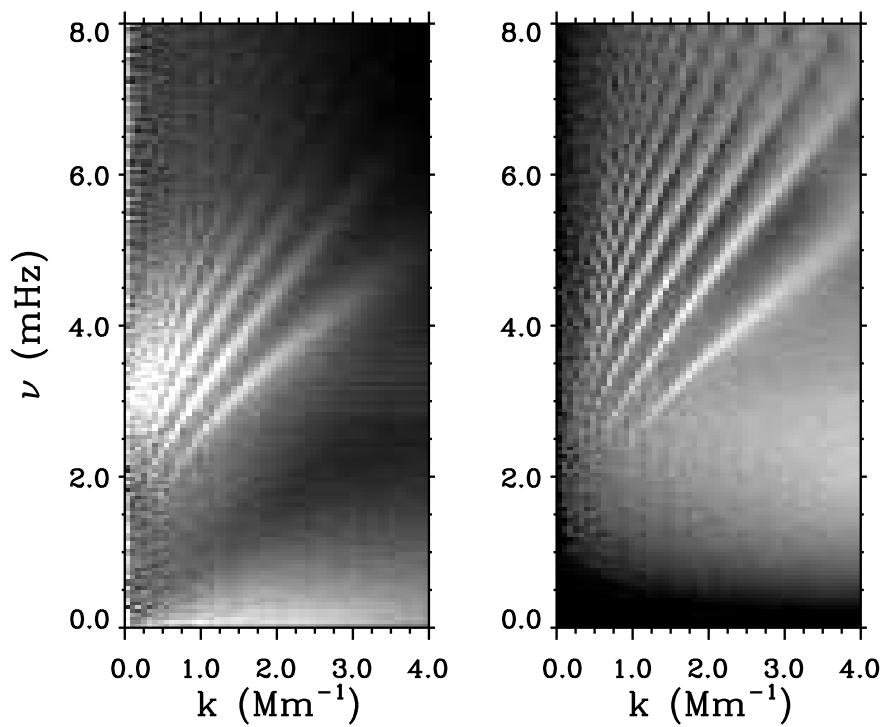} \caption{Power spectrum diagrams
from MDI high resolution observations ({\it left}) and Model~I of
our simulations ({\it right}). Both power spectra are computed using
the same time duration and after the simulated data are binned down
to the same spatial resolution as the observed data.} \label{fg3}
\end{figure}

The acoustic power spectrum ($k$-$\omega$ diagram) of the simulated
wave field is shown in Figure~\ref{fg3}. We see a good agreement
with the observed power spectrum in terms of the shape and locations
of power ridges, yet the simulated wave field has more power in
high-frequency region. It is not clear whether such power excess
really exists and instrumentally filtered out during observations or
it is an artifact of the numerical modeling of the wave field. The
realistic non-linear simulations of solar convection also show a
power excess in the $k$-$\omega$ diagram at higher frequencies
\citep{Georgobiani2007}. For the present study, the high-frequency
power excess is unimportant, because for the time-distance
helioseismology analysis, we only select the frequency band of 3 --
4 mHz for analysis, and in this range, the simulated acoustic power
is similar to the observations.

\section{Results}
\subsection{Time-Distance Analysis}
To perform the time-distance helioseismology analysis of the
simulated data we followed the procedure described in
\citet{Zhao2001}. For all our models, we select the annuli with
radii of $6.2 - 11.2$ Mm, $8.7 - 14.5$ Mm, and $14.5 - 19.4$ Mm to
obtain the mean travel times, $\tau_\mathrm{mean}^{(i)}$, and the
travel time differences, $\tau_\mathrm{diff}^{(i)}$, (index $i$
marks model number, including the reference Model~I), which are
respectively averages and differences of outgoing and ingoing travel
times in the time-distance center-annulus measurement scheme. A
phase-speed filter is applied in each case to select only waves in a
narrow phase speed interval. To study the effects caused by
wavefield non-uniformity, we calculate the differences of
$\delta\tau_\mathrm{mean}^{(i)} = \tau_\mathrm{mean}^{(i)} -
\tau_\mathrm{mean}^\mathrm{(I)}$ and $\delta\tau_\mathrm{diff}^{(i)}
= \tau_\mathrm{diff}^{(i)}-\tau_\mathrm{diff}^\mathrm{(I)}$ for the
analysis.

\begin{figure}
\epsscale{0.9} \plotone{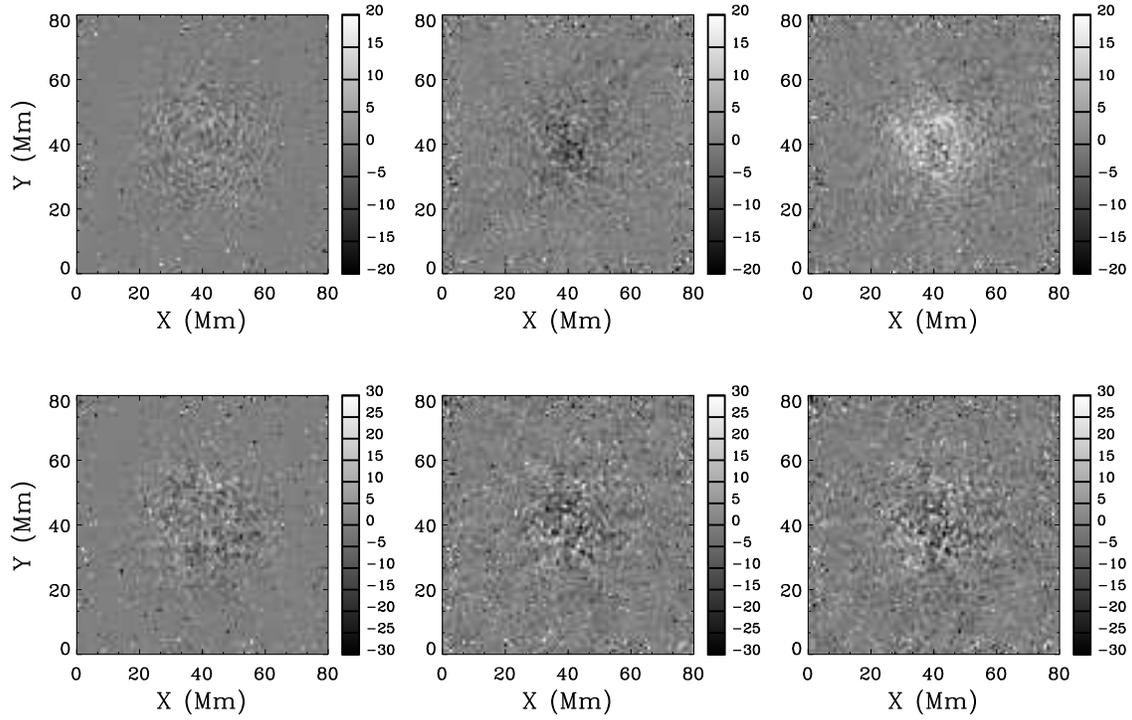} \caption{Maps for mean travel times
({\it upper row}) and travel time differences ({\it bottom row})
relative to the Model~I. For both rows, from the left column to the
right are for Model~Ia, Model~II, and Model~III, respectively.
Annulus radii used for time-distance measurements are from 8.7 to
14.5 Mm.} \label{fg4}
\end{figure}

In Figure~\ref{fg4}, we show the maps of the mean travel time
perturbations, $\delta\tau_\mathrm{mean}^{(i)}$, the travel time
differences, $\delta\tau_\mathrm{diff}^{(i)}$, for all three models.
Although the background model of Models~Ia and II are the same with
the reference Model~I, we can see systematic shifts of mean travel
times inside the masked regions.

For better understanding the results, it is useful to compare the
profiles of the travel time deviations, azimuthally averaged around
the sunspot center for both $\delta\tau_\mathrm{mean}^{(i)}$ and
$\delta\tau_\mathrm{diff}^{(i)}$, as shown in the middle row of
Figure~\ref{fg5}. Obviously, the mean travel time shifts,
$\delta\tau_\mathrm{mean}$, are significantly larger in Model~II
than in Model~Ia, although both have exactly the same background
model and exactly the same oscillation amplitude reduction in
wavefields. Expectedly, Model~III shows mostly positive travel time
shifts in contrast with the other two models, and this is obviously
due to the negative sound-speed perturbation to the background model
in a shallow subsurface region. One would expect this positive time
shift would increase significantly if there is no such effect that
causes the time deficit in Model~II, however, it is not immediately
clear whether this is the case, or if it is, how much it would
increase. For the travel-time difference,
$\delta\tau_\mathrm{diff}$, the shifts for all three models are
quite small, within an order of 2 sec, substantially smaller than
the measured travel time shifts from a real sunspot data
\citep{Zhao2001}.

The azimuthally averaged $\delta\tau_\mathrm{mean}^{(i)}$ and
$\delta\tau_\mathrm{diff}^{(i)}$ for the other two annuli
measurements are also presented in Figure~\ref{fg5}. For the shorter
travel distances, both Models~Ia and II show stronger travel time
deficits in the mean travel time measurements compared the the
intermediate travel distance case, up to approximately 15 sec for
Model~II. However, Model~III still displays mostly a positive sign,
although it displays some dips in the central area where one would
expect a stronger positive shift because of the larger negative
sound-speed perturbation there. Again,
$\delta\tau_\mathrm{diff}^{(i)}$ does not show any significant time
shifts for all models. For the larger annulus radius, Models~Ia and
II do not display significant time shifts, but Model~III displays a
significant negative time shift because for this set of measurement,
waves reach the depth of a large positive sound-speed perturbation.
For this annulus measurement, Models~II and III display an order of
magnitude of 5 sec travel time shifts in $\delta\tau_\mathrm{diff}$,
larger than those of shorter annuli measurements, but still
significantly smaller than the shifts in the real sunspot
measurements.

\subsection{Power Correction}
Based on their artificial tests with the quiet Sun data, in order to
remove the measured travel time shifts caused by the oscillation
amplitude reductions, \citet{Rajaguru2006} have suggested to make
corrections for these areas by enlarging the observed oscillation
amplitude in active regions. This procedure is just reverse to the
artificial masking. It obviously works if the power reduction is
caused by artificial masking, like in Model~Ia. However, it is
useful to examine whether this works for the oscillation power
reductions that are not caused by surface masking, but physical
mechanisms, such as the reduction in the excitation power (like in
Models~II and III).

For each model (Ia, II, and III) we calculate the average amplitude
profile and normalized the wavefield by using this profile
(procedure of unmasking the wavefield), making the oscillation power
nearly uniform over the whole box. The same time-distance analysis
is performed like in \S3.1, and azimuthally averaged curves are
displayed in Figure~\ref{fg6}. It can be clearly seen that, as
expected, this power correction removes all travel time shifts in
both $\delta\tau_\mathrm{mean}$ and $\delta\tau_\mathrm{diff}$ for
all annuli measurements for Model~Ia. For
$\delta\tau_\mathrm{mean}$, for the two shorter annuli measurements,
the correction slightly lifts both Models~II and III without
changing signs of the profiles, and for the longest annulus, the
correction does not change much the measurements. For
$\delta\tau_\mathrm{diff}$, the correction changes the profiles of
Models~II and III for all annuli, but still, the travel time shifts
are within 5 sec or so.

\section{Discussion and Conclusion}
The explanation of the fact that the acoustic travel times depend on
the non-uniformness of the wave field amplitude or non-uniform
distribution of the source strength is related to the definition of
travel times in helioseismology, which have to deal with the
stochastic randomly excited oscillations, rather than isolated point
sources. The travel time of a wave packet traveling between two
points  on the surface is defined not as a local physical quantity,
which can be explicitly computed from the background model, but
rather in an ``observational" way as a parameter which is obtained
from fitting of cross-correlation of the oscillation signals by the
Gabor wavelet. Thus, it is very important to investigate the effect
of non-uniform distribution of acoustic sources, damping and other
causes of the non-uniform wave field distribution on the Sun. We
have presented the results for some of these effects by using
numerical 3D simulations of acoustic wave propagation in various
solar models.

We have found that the source masking for horizontally uniform
background model (dashed curves) may cause a systematic negative
shift of about 8--13 seconds in the mean travel times for short
distances (annuli with radii smaller than 14.5 Mm). Such travel time
shift may cause underestimation of the sound speed perturbation in
the shallow (1--2 Mm deep) subsurface layers. For larger distances,
the contribution to the mean travel time shift becomes negligible.
On the contrary, the shift of the travel time differences (due to
the non-uniform distribution of the acoustic sources) is negligible
for short distances and has a value about $-5$ sec for the largest
distance used in our experiments. This is much smaller than
perturbations of the travel-time differences observed in real
sunspots.

\begin{figure}
\plotone{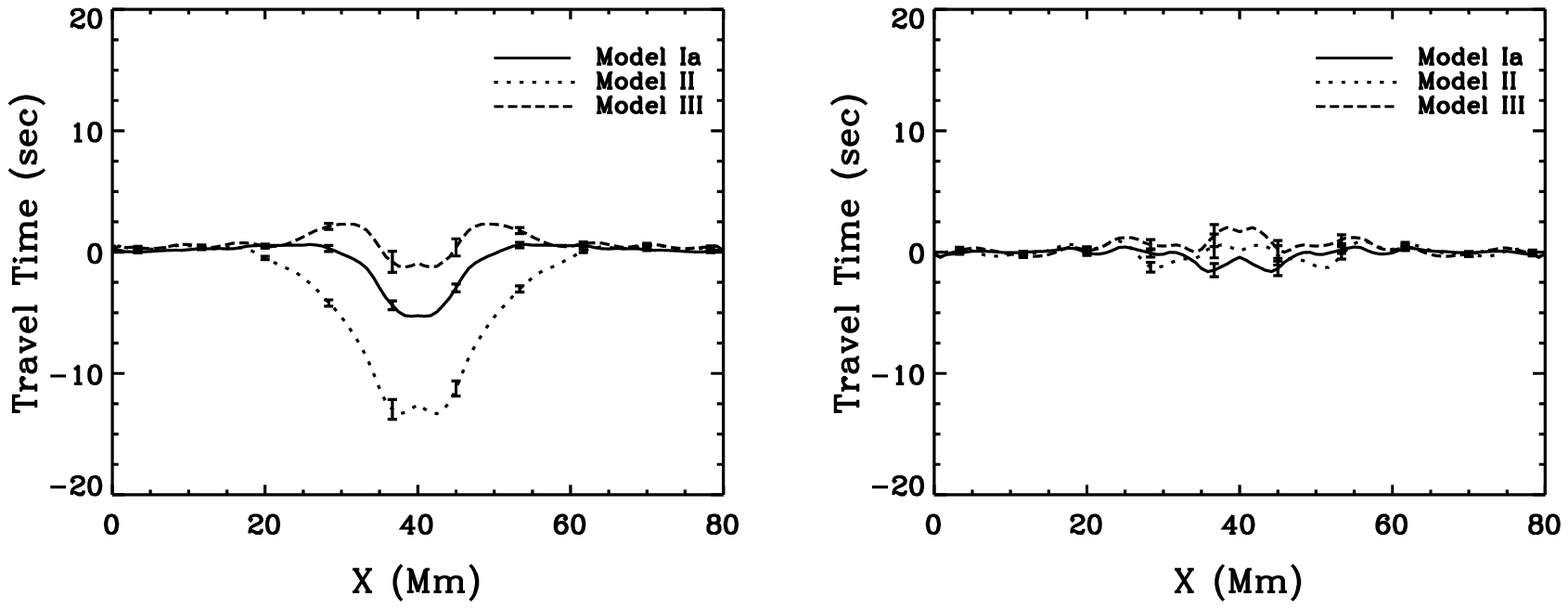} \vskip 5mm \plotone{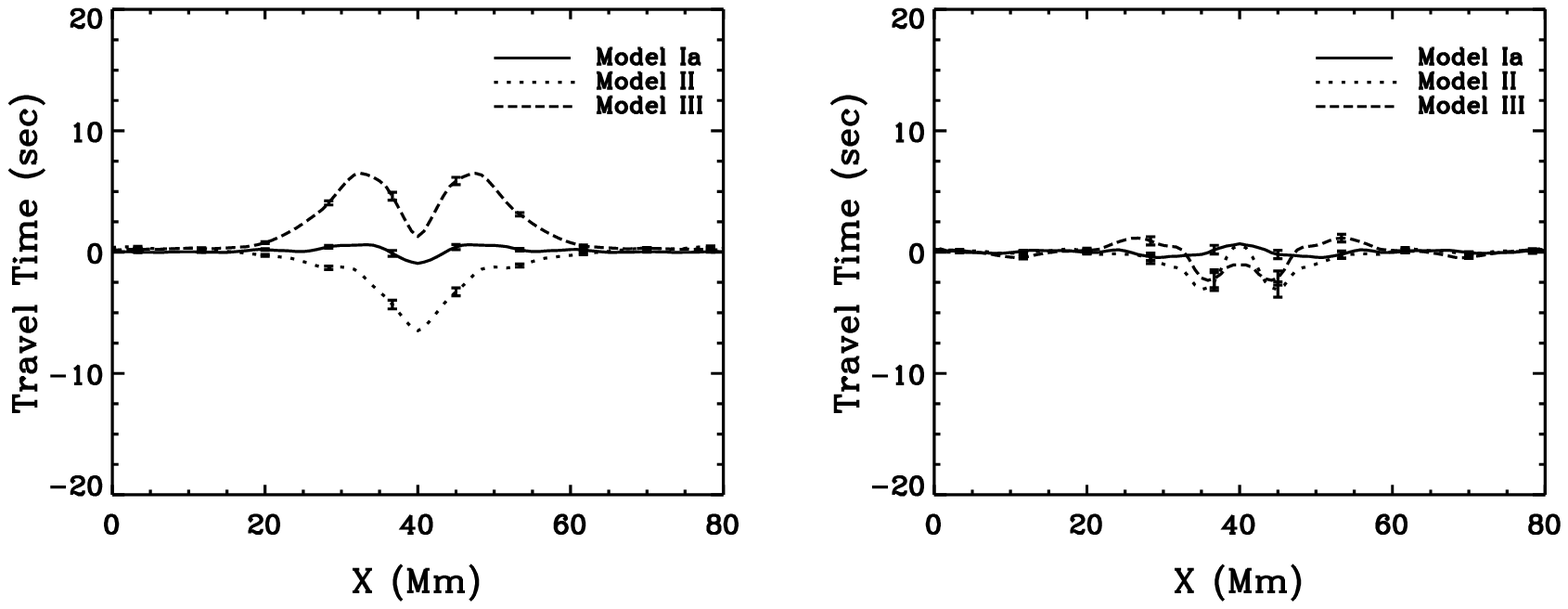} \vskip 5mm
\plotone{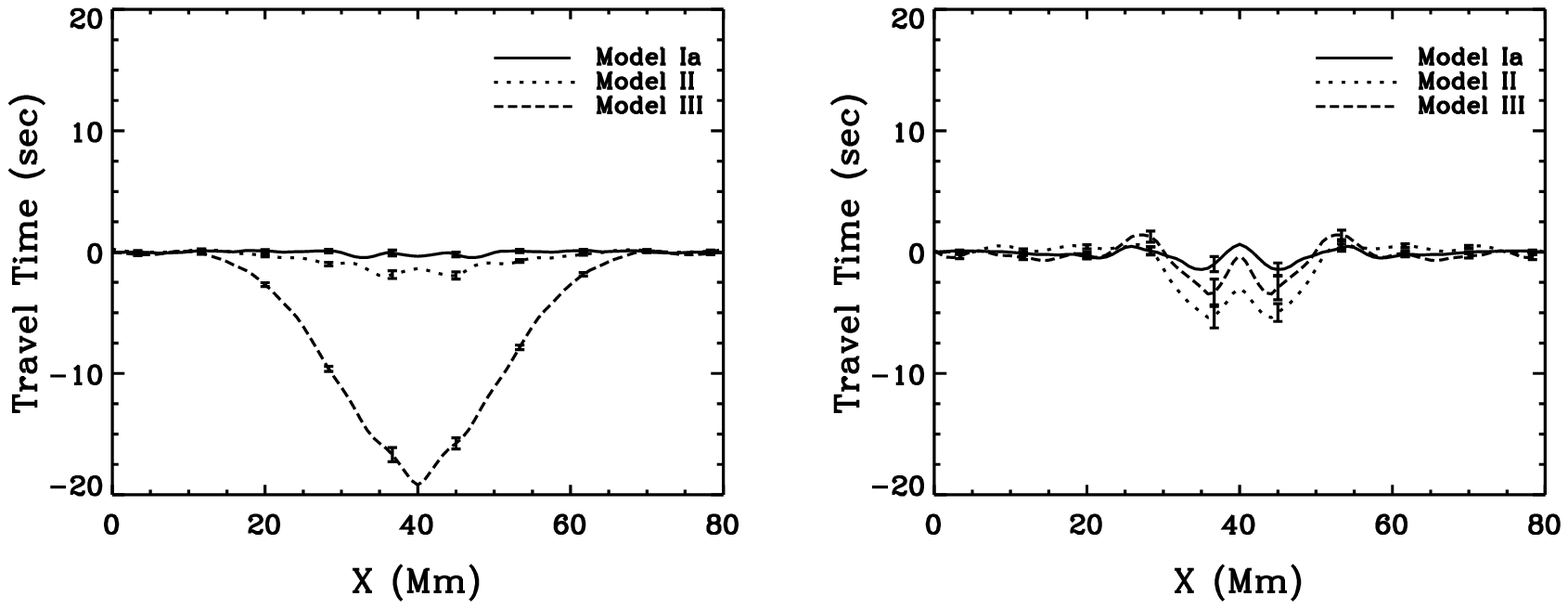} \caption{Mean travel times ({\it left panels}) and
travel time differences ({\it right panels}) azimuthally averaged
from maps like shown in Figure~\ref{fg4} for different annulus radii
of (from above to bottom) $6.2-11.2$ Mm, $8.7 - 14.5$ Mm, and $14.5
- 19.4$ Mm.} \label{fg5}
\end{figure}

\begin{figure}
\plotone{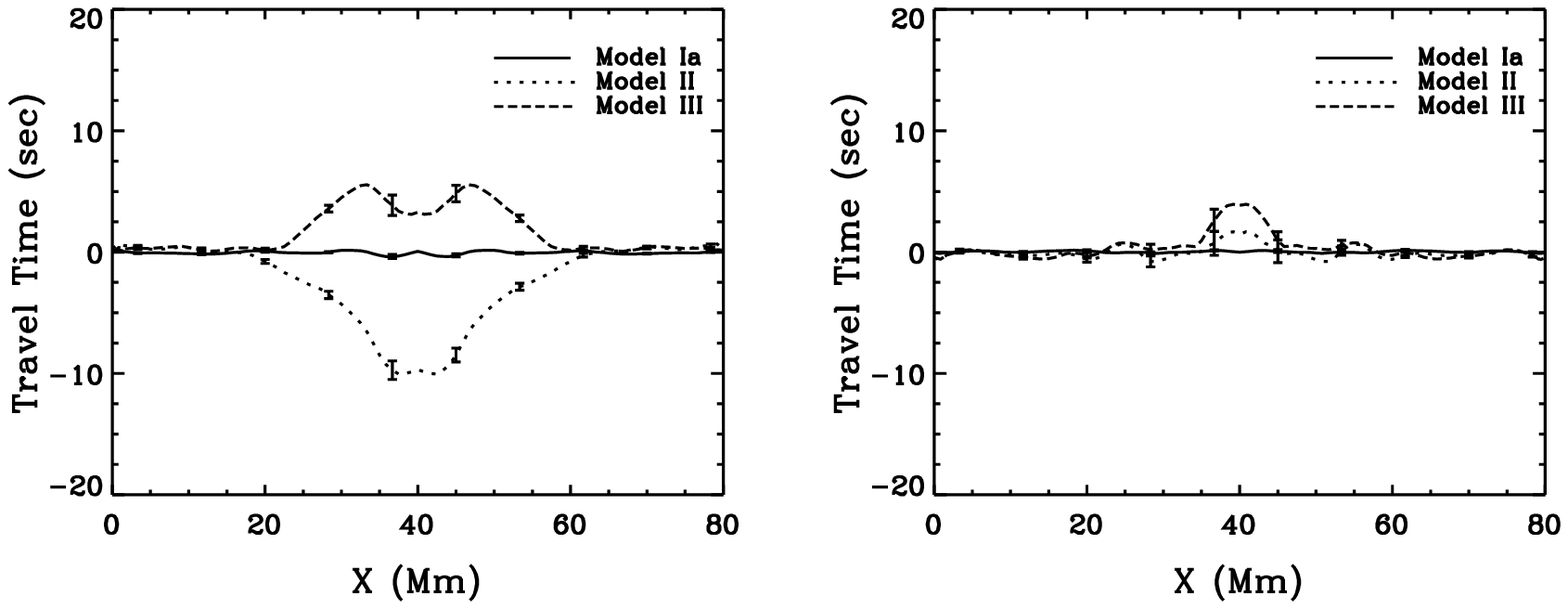}
\vskip 5mm
\plotone{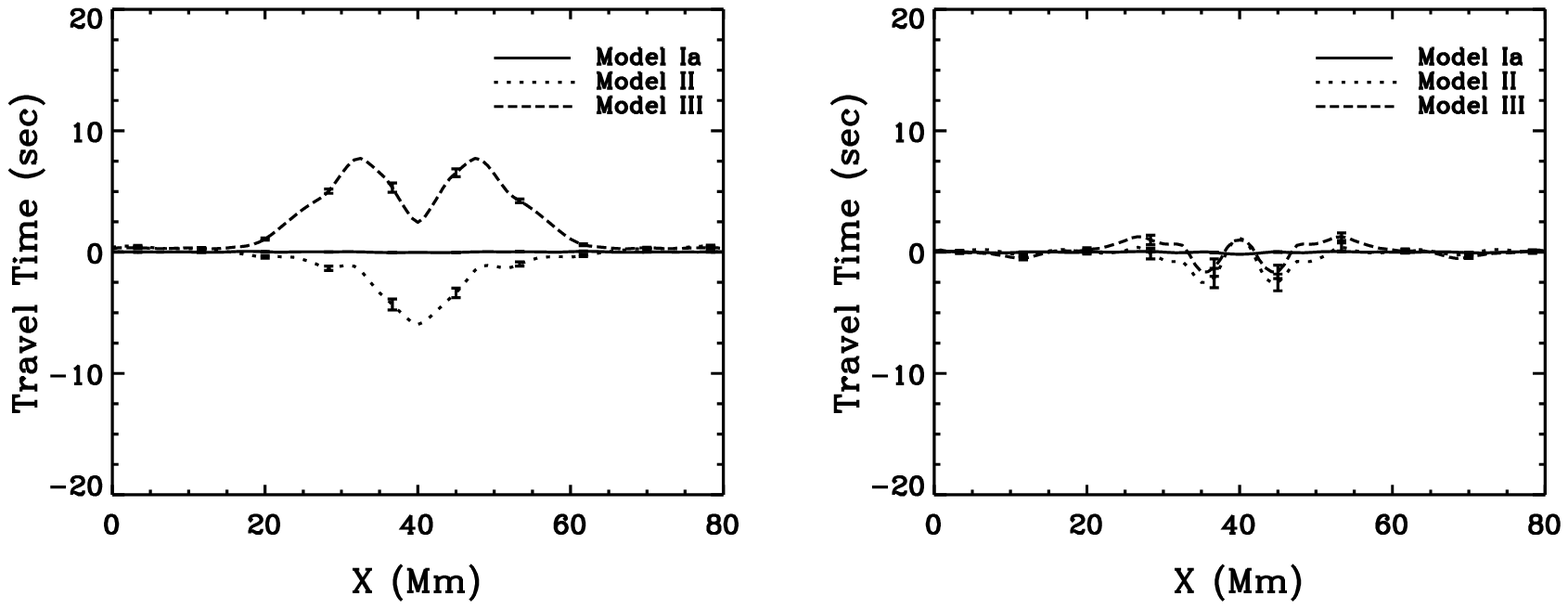}
\vskip 5mm
\plotone{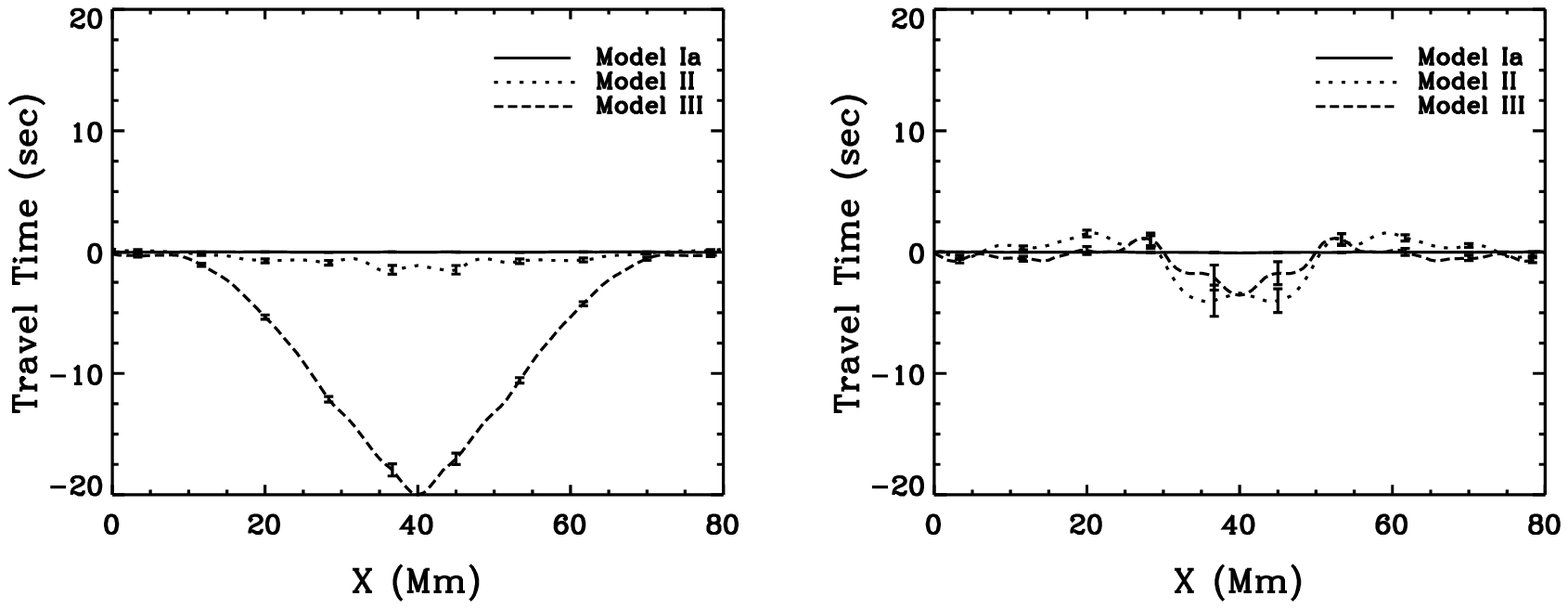}
\caption{Same as Figure~\ref{fg5}, but after the power corrections.}
\label{fg6}
\end{figure}

The results of our experiments are different from  a similar work by
\citet{Hanasoge2007}, where authors report significant disbalance
between ingoing and outgoing travel times (about $-5$ s for distance
of 6.2 Mm and about $-15$ s for distance of 24.35 Mm), and suggested
that at the large distances the false travel time difference signal
caused by non-uniform distribution of sources may be misinterpreted
as a result of subsurface flows. However, it is quite clear that, as
shown in the first two annuli measurements of Figure~\ref{fg5}, the
oscillation power deficit due to the source masking may have greatly
reduced the travel time shifts measured in Model~III, which means
that if doing inversions, the inverted sound-speed profile would be
greatly underestimated. This suggests that the sound-speed profile
under sunspots obtained by \citet{Kosovichev2000} got the correct
sign but might be underestimated. For the flow fields, this masking
effect might cause some systematic velocity errors, but only of a
very small magnitude.

In addition, our experiments show that the amplitude reduction is
caused by the weaker oscillation sources in sunspots cannot be
corrected by a simple normalization procedure. This imposes us a
difficult task on how to retrieve accurately the sound-speed
profiles beneath sunspots, and improve the time-distance
helioseismology inferences.

\clearpage

\clearpage

\clearpage

\clearpage

\clearpage

\clearpage

\end{document}